\documentclass[aps,prl,twocolumn,showpacs,superscriptaddress,floatfix]{revtex4-1}  
\usepackage{ucs}
\usepackage{natbib}
\usepackage{graphicx}  
\usepackage{bm}        
\usepackage{amssymb}   
\usepackage{amsmath}
\usepackage{color}
\usepackage{CJK}
\usepackage{times}
\usepackage{verbatim}
\usepackage{mathrsfs}
\usepackage{hyperref}
\usepackage{ulem}
\hypersetup{colorlinks = True, urlcolor=blue, linkcolor=blue, citecolor=blue}

\newcommand{\sig}[2]{\sigma^{#1}_{#2}}
\newcommand{\tu}[2]{\tau^{#1}_{#2}}

\usepackage{cancel}
\usepackage{comment}
\begin{document}

\title{Spin- and Flux-gap Renormalization in the Random Kitaev Spin Ladder}
\author{Wen-Han Kao}
\email{kao00018@umn.edu }
\affiliation{School of Physics and Astronomy, University of Minnesota, Minneapolis, MN 55455, USA}

\author{Natalia B. Perkins} 
\email{nperkins@umn.edu}
\affiliation{School of Physics and Astronomy, University of Minnesota, Minneapolis, MN 55455, USA}

\date{\today}
\begin{abstract}
We study the Kitaev spin ladder with random couplings by using the real-space renormalization group technique. This model is the minimum model in Kitaev systems that has conserved plaquette fluxes, and its quasi-one-dimensional geometry makes it possible to study the strong-disorder fixed points for both spin- and flux- excitation gaps. In the Ising limit, the behavior of the spin gap is consistent with the familiar random transverse-field Ising chain, but the flux gap is dominated by the y-coupling. In the XX limit, while the x- and y-couplings are renormalized simultaneously, the z-couplings are not renormalized drastically and lead to non-universal disorder criticality at low-energy scales.
\end{abstract}
\maketitle

\textit{Introduction}.---Quantum spin liquid (QSL), an exotic magnetic phase with fractionalized spin excitations and intricate entanglement structure, has been pursued both theoretically and experimentally since its first proposal by Anderson in 1973~\cite{Anderson1973}. Theoretical models and candidate materials with strong geometrical or exchange frustration are expected to greatly reduce the ordering temperature and reveal the quantum fluctuations. However, the presence of residual interactions in real systems usually leads to magnetic ordering and shatters the hope for finding QSL. Nevertheless, various compounds were discovered with no magnetic ordering even down to the lowest measurable temperature, and commonly the quenched randomness was found to serve as a potential cause of the sustaining disordered phase and intriguing dynamics of low-energy degrees of freedoms ~\cite{Freedman2010,Riedl2019, Yamaguchi2017,Murayama2020,Do2020}. Therefore, the competition between quantum fluctuations and randomness raises a critical question about the true nature of the low-energy phase in those materials. For example, the peculiar low-energy excitations  found in a second-generation Kitaev materials~\cite{Kitagawa2018spin,Trebst2022,Bahrami2022review} may be ascribable to spin fractionalization in weakly disordered QSL~\cite{Knolle2019,Kao2021vacancy,Kao2021localization}, but it may also relate to the random singlet (RS) phase in strongly disordered magnets~\cite{kimchi2018scaling}. The RS phase was first proposed in the studies of the random spin chains by a real-space perturbation scheme called the strong-disorder renormalization group (SDRG)~\cite{Ma1979, Dasgupta1980,Fisher1994}. This method has been shown to be asymptotically exact if the random distribution broadens infinitely through the iteration, and the ground-state wavefunction of the RS is a product state of singlets with a variety of distances. In this regard, the RS phase is not a QSL despite having quantum critical features.

Even though the strong-disorder phenomenology for Kitaev systems has been discussed in different scenarios~\cite{kimchi2018scaling,Sanyal2021}, the direct application of SDRG is still lacking and the fate of flux degrees of freedom under disorder remains an open question. Previous studies of SDRG on two-dimensional models have shown that the connectivity on each site grows drastically in the iteration~\cite{Motrunich2000,Lin2003,Kovacs2010,Kovacs2011}, and thus it is presumed that any properties derived from elementary plaquettes of the lattice, such as geometrical frustration and plaquette fluxes, will be destroyed by strong disorder.


In this Letter, we propose a quasi-one-dimensional spin ladder with plaquette fluxes that survive the strong disorder which can be accounted for  by the  SDRG approach extended to study the disorder criticality not only in the spins but also in flux sectors. 
  Our numerical SDRG results show that the spin- and flux-gap energy scale can either be a universal function following the SDRG fixed-point distribution, or a non-universal one affected by the initial random distributions.
Here we present our main  findings and  defer full technical details  of the extended SDRG methodology  to a forthcoming publication~\cite{Kaounpublished}.

\begin{figure}
     \includegraphics[width=1.0\columnwidth]{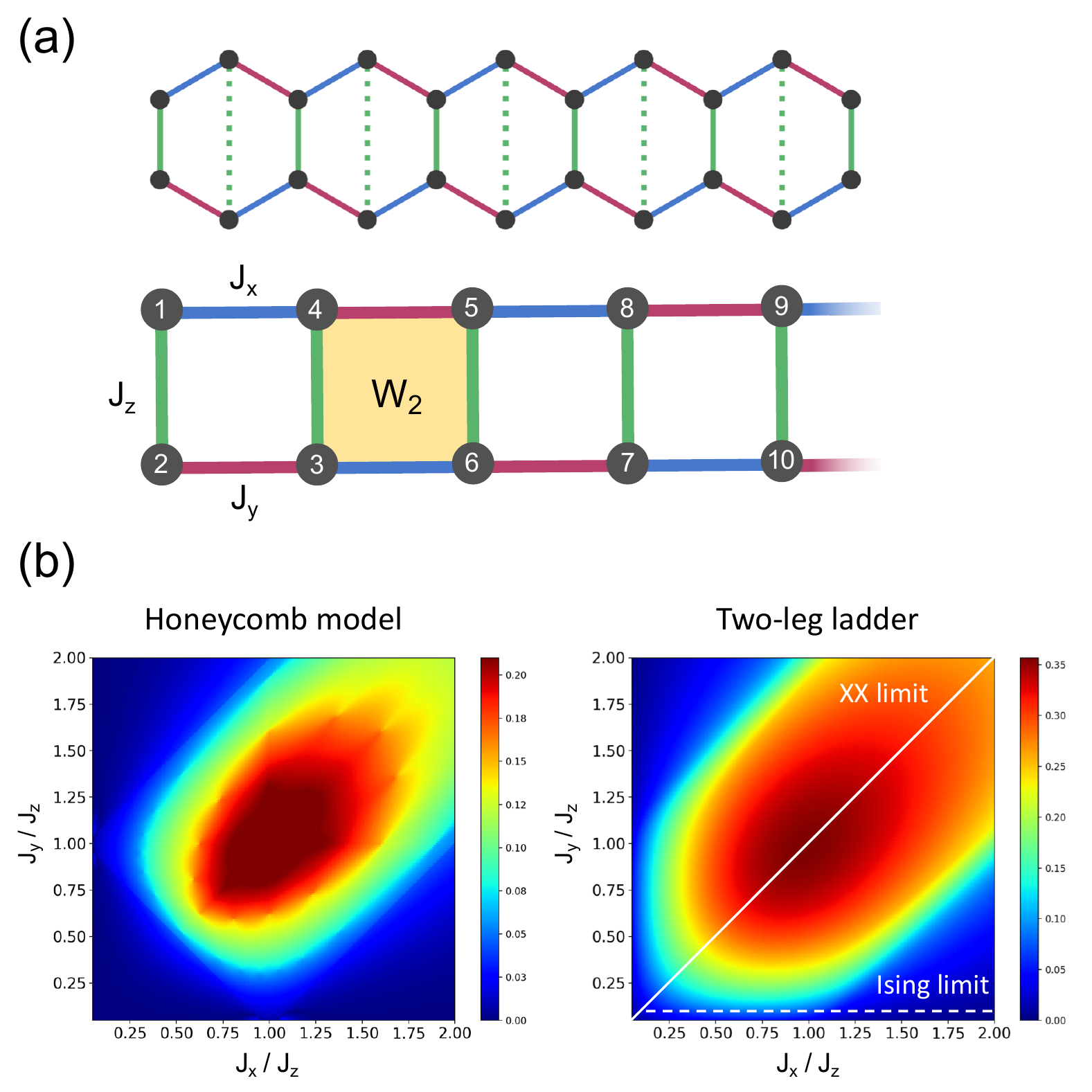}
     \caption{\label{fig:Figure1}(a) Kitaev spin ladder derived from one-row of the honeycomb lattice. The shaded plaquette denotes the conserved flux operator $W_{2} = -\sig{y}{3}\sig{x}{4}\sig{x}{5}\sig{y}{6}$. (b) Lowest flux excitation gap for pure Kitaev honeycomb model and two-leg ladder. The white solid (dashed) line denotes the XX limit (Ising limit) of the model considered in this work. In the Ising limit, the small but non-zero $J_{y}$ is required in order to have finite flux gaps.}
\end{figure}

\begin{figure*}
     \includegraphics[width=1.0\textwidth]{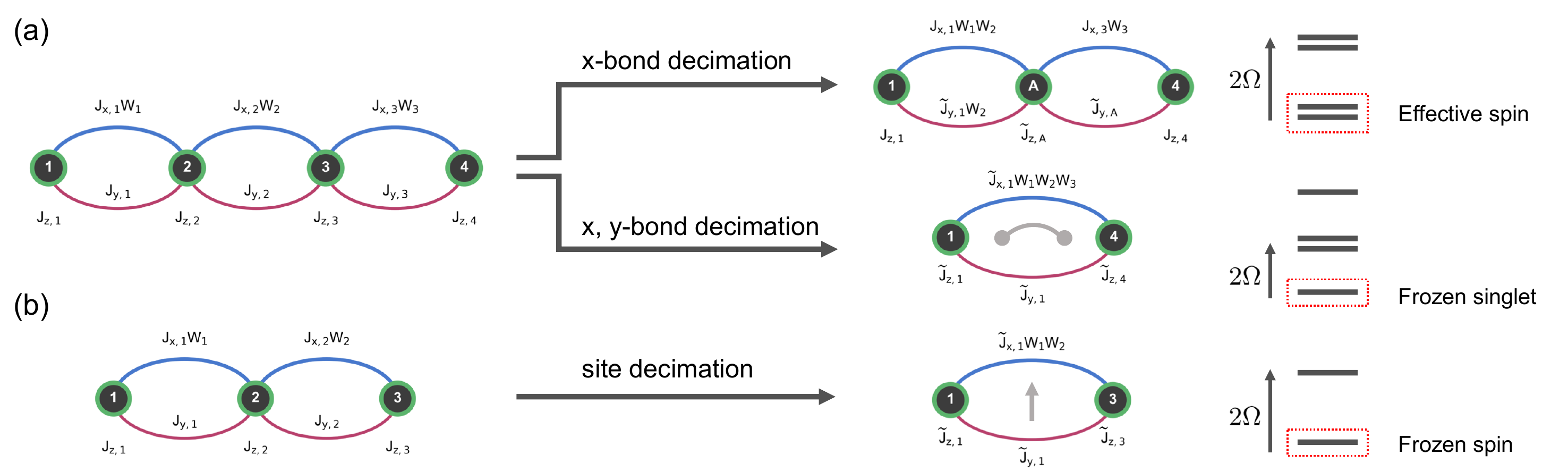}
     \caption{\label{fig:Figure2} Decimation rules for the Kitaev spin ladder in the Ising  and XX limits. (a) Ising limit (upper panel): The x-bond decimation with largest energy scale $\Omega = J_{x,2}$. The upper doublet formed by site 2 and site 3 is integrated out and the lower doublet forms an effective spin-$1/2$  site  A with renormalized couplings and  an on-site field.   XX limit (lower panel):  x- and y-couplings on the same bond are equal, $J_{x,j}=J_{y,j}\equiv J_j$, so they are simultaneously decimated with the largest energy scale $\Omega = J_{2}$.  Spins on site 2 and 3 form a local ground state with a large decimation gap $2\Omega$, and spins on site 1 and  4  couple by a  renormalized interaction with a new flux variable.
     (b) The site decimation with largest energy scale $\Omega = J_{z,2}$. Site 2 is aligned by the transverse field and the neighboring sites are coupled by renormalized couplings.}
\end{figure*}

\textit{Strong-disorder renormalization group}.--- The main idea of SDRG is to successively decimate out the larger-energy (shorter-time) degrees of freedom in real space and scrutinize the flow of random distributions throughout the iteration. The low-energy physics, phase transitions, and corresponding critical exponents can be extracted from the fixed-point solutions. Moreover, for simple one-dimensional models such as random Heisenberg chain and random transverse-field Ising chain (RTIC), the analytical solution of the RG flow equation can be obtained even in the off-critical Griffiths region away from infinite-disorder fixed point (IDFP)~\cite{Igloi2002}. In more complicated models, numerical SDRG was applied extensively in order to extract the low-energy physics governed by strong-disorder fluctuations. Comprehensive reviews of SDRG can be found in Ref.~\cite{Kovacs2011,Igloi2018}.

\textit{Model}.---We consider a spin-$1/2$ two-leg Kitaev spin ladder with alternating $x-$ and $y-$ Ising couplings on the legs and $z-$couplings on the rungs as in Fig.~\ref{fig:Figure1} (a). The sites are labeled as an one-dimensional chain with third-neighbor $x$-couplings:
\begin{equation}
    \mathcal{H} = -\sum_{l=1}^{N/2}\left[ J_{x}\sig{x}{2l-1}\sig{x}{2l+2}+J_{y}\sig{y}{2l}\sig{y}{2l+1}+J_{z}\sig{z}{2l-1}\sig{z}{2l} \right].
\end{equation}
 This can be seen as a minimum model reduced from the Kitaev honeycomb model~\cite{Kitaev2006} that still has conserved flux operators $W_{l} = -\sig{y}{2l-1}\sig{x}{2l}\sig{x}{2l+1}\sig{y}{2l+2}$ on each plaquette. We apply the duality transformation~\cite{Feng2007}
\begin{equation}
    \sig{z}{j} = \prod_{k=j}^{N}\tu{z}{k}, \quad \sig{y}{j} = \tu{y}{j-1}\tu{y}{j}, 
\end{equation}
such that the effective Hamiltonian (excluding the boundary terms) becomes a transverse-field XY pseudospin chain:

\begin{equation}
    \mathcal{H} = -\sum_{j}^{L}\left[ J_{x}W_{j}\tu{x}{j}\tu{x}{j+1}+J_{y}\tu{y}{j}\tu{y}{j+1}+J_{z}\tu{z}{j} \right].
\end{equation}

Note that we relabel the indices since the couplings and fields are present only on odd sites ($2l-1$) such that the length of the chain is $L=N/2$. In addition, the operators of even sites ($2l$) form a set of local conserved quantities $W_{l}=\tu{y}{2l-2}\tu{z}{2l}\tu{y}{2l+2}$ with eigenvalues $\pm 1$. In fact, this is precisely the plaquette flux operator of the original spin model. In this dual Hamiltonian, the fluxes are present explicitly so that we can separate flux sectors by different sets of $W$ eigenvalues. According to the Lieb's theorem~\cite{Lieb1994}, the ground-state sector of the pure Kitaev ladder is in the $\pi$-flux phase where all $W=-1$. With open boundary condition, the flux can be singly excited from $-1$ to $+1$. The lowest flux excitation energy is plotted in Fig.~\ref{fig:Figure1} (b) and it resembles the flux-gap phase diagram of the Kitaev honeycomb model in Fig.~\ref{fig:Figure1} (a) ~\cite{Motome2019}.

In the random model, all the couplings and fields are quenched-disordered and given by arbitrary random distributions $R(J_{x})$, $Q(J_{y})$, and $P(J_{z})$.

\textit{XX limit}.---In this limit, the $x$- and $y$-couplings on the same bond are equal $J_{x,j}=J_{y,j}\equiv J_{j}$ and thus only two independent distributions $R(J)$ and $P(J_{z})$ are considered. In the bond decimation scheme (Fig.~\ref{fig:Figure2} (a), lower panel), we derive the RG rule from the four-site example with the largest energy scale $\Omega = J_{2}$, and treat the rest of terms as perturbation.  The two spins in the middle are then frozen in their local ground state with a large decimation gap $2\Omega$, similar to the formation of random singlet in random Heisenberg chain. Spins on site 1 and  4 then form a renormalized coupling with a new flux variable:
\begin{align}
    \tilde{J} = \frac{J_{1}J_{3}}{\Omega}, \quad \tilde{W} = W_{1}W_{2}W_{3}.
\end{align}
With $W_{2}=\pm 1$, $J_{z,2}$ and $J_{z,3}$ will contribute to the different constant energy shift in the perturbation theory, and this gives rise to the flux gap
\begin{align}\label{Eq:bond_flux_gap}
\Delta E_{f}(W_{2}) = \frac{J_{z,2}J_{z,3}}{\Omega}.
\end{align}

In the site decimation scheme, we consider a three-site example with $\Omega = J_{z,2}$ (Fig.~\ref{fig:Figure2} (b)). The strongest field tends to align the spin on site 2 and renormalized couplings are formed between site 1 and site 3 as $\tilde{J} = J_{1}J_{2}/\Omega$ with $\tilde{W} = W_{1}W_{2}$.
However, in site decimation the neighboring transverse fields are also modified and become flux-dependent:
\begin{align}\label{Eq:site_flux_gap}
    \tilde{J}_{z,1} = J_{z,1}-\frac{J_{1}^{2}}{\Omega}W_{1}, \quad \tilde{J}_{z,3} = J_{z,3}-\frac{J_{2}^{2}}{\Omega}W_{2}.
\end{align}
In later decimation steps, if $\tilde{J}_{z,1}$ or $\tilde{J}_{z,3}$ happen to be the new $\Omega$ of the system, this flux-dependent term will become the flux gap for $W_{1}$ or $W_{2}$. One important observation from the decimation rules is that the couplings are renormalized much faster than the fields, such that the site decimation becomes dominant in the low-energy limit~\cite{Kaounpublished}. Numerically, we implemented SDRG in a similar way to the studies of random Heisenberg chain and RTIC for the spins, and additionally calculate the local flux gaps from Eq.~(\ref{Eq:bond_flux_gap}) and Eq.~(\ref{Eq:site_flux_gap}) for each decimation step, if applicable.

\textit{Ising limit}.---In this limit, we consider the scale of $J_{y}$ to be much smaller than $J_{x}$ and $J_{z}$, such that $x$-bond and site decimations are dominant in SDRG. Therefore, the fixed-point solution of RTIC can be applied for $J_{x}$ and $J_{z}$. Even though $J_{y}$ is considered to be very small, it is required to be finite in order to have non-zero flux excitation energies. In the bond-decimation scheme ($\Omega = J_{x,2}$), the strongest Ising coupling leads to a low-energy doublet which is kept as a renormalized spin, as shown in upper panel of Fig.~\ref{fig:Figure2} (a). The low-energy effective Hamiltonian can be rigorously derived via the Schrieffer-Wolff transformation~\cite{Altman2014, Altman2014PRX}, and the effective transverse field on this renormalized spin is~\cite{Kaounpublished}
\begin{align}\label{Eq:JzA}
    \tilde{J}_{z,A} = J_{y,2}W_{2}-\frac{J_{z,2}J_{z,3}}{\Omega}.
\end{align}
The site decimation rule is derived from the three-site example and is similar to that in the XX limit.

Since the decimation rules in the Ising limit are more complicated than in the XX limit, we executed the SDRG calculations on the $\pi-$flux and all the one-flux excitation sectors for each random realization. The flux gaps are then computed from the energy difference between different sectors, instead of estimated from the decimation rules. The validity of the method is verified by comparing to the small-sized results in exact diagonalization.

\begin{figure}
     \includegraphics[width=0.9\columnwidth]{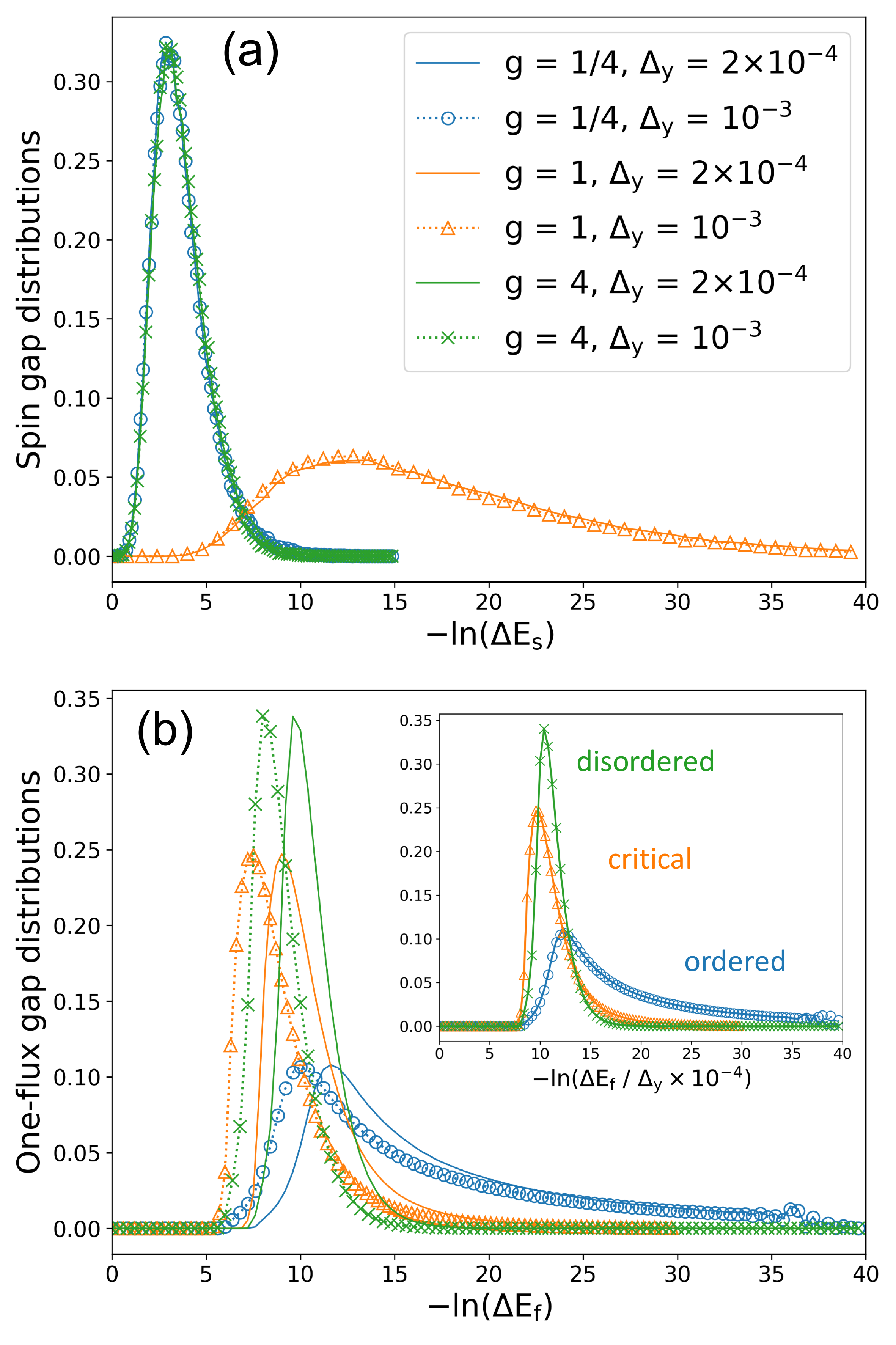}
     \caption{\label{fig:Figure3} Distributions of spin and flux gaps of different phases for various small $\Delta_{y}$ in the Ising limit. (a) The spin gaps for one-flux sectors. The distributions of the first excitation gap are shown for disordered (g=4) and critical phase (g=1), and the second gap is chosen for ordered phase (g=1/4). (b) Distribution of the one-flux excitation gaps. The inset shows that the $\Delta_{y}$ dependence can be removed by dividing $\Delta_{y}$ for each gap. All data are obtained from the numerical SDRG of $10^{5}$ random samples of $L=128$ chain.}
\end{figure}

In the numerical SDRG calculation, we consider different phases according to the initial uniform distributions of the couplings with different widths:
\begin{align}
J_{x}\in [0,\Delta_{x}], \quad J_{y}\in [0,\Delta_{y}], \quad J_{z}\in [0,\Delta_{z}].
\end{align}
For the critical phase we set $g = \Delta_{z}/\Delta_{x}=1$, and for the off-critical (Griffiths) phase we take $g= 1/4$ or $4$. In all cases we set $\Delta_{y}$ to be much smaller such that y-bond decimation happens rarely.
In Fig.~\ref{fig:Figure3}, we show the distributions of spin and flux gaps from all the one-flux excitation sectors. Note that in the disordered ($g=4$) and critical ($g=1$) phases we collect the first spin gap while in the ordered ($g=1/4$) phase we use the second spin gap, due to the fact that in the numerical simulations the first spin gap corresponds to finite-size effect of degenerate ground states in the ordered phase~\cite{Igloi2005,Extreme2006}. For the spin gaps, the ordered and disordered phases have identical distribution because of their duality with respect to the critical point. Also, as shown in Fig.~\ref{fig:Figure3} (a), the spin-gap distributions are almost unchanged by tuning the width $\Delta_{y}$. However, in the flux-gap distributions  shown in Fig.~\ref{fig:Figure3} (b), a clear shift is observed for different $\Delta_{y}$ in all three phases. This suggests that while the spin gaps are functions of $J_{x}$ and $J_{z}$ and almost independent of $J_{y}$ values, the flux gaps in general have linear dependence in $J_{y}$. 

\begin{figure*}
     \includegraphics[width=1.0\textwidth]{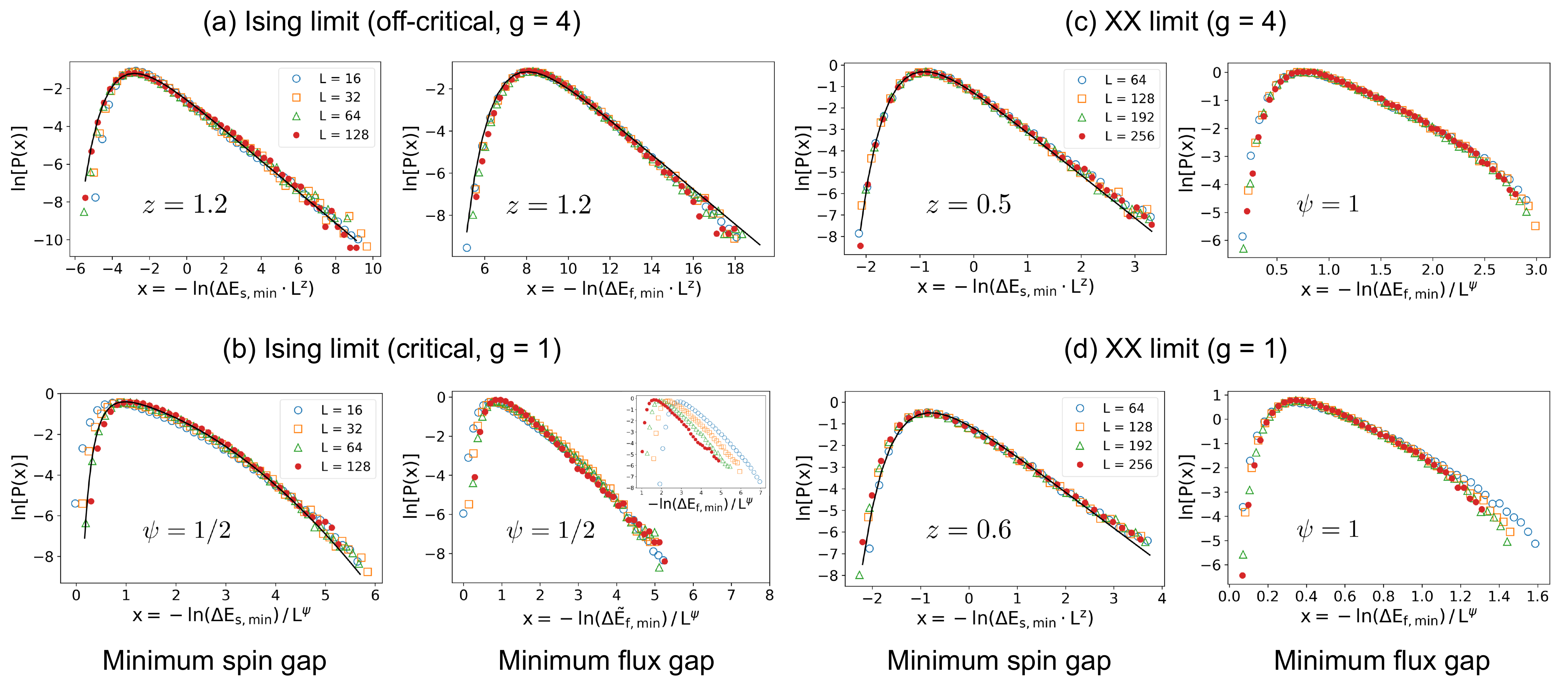}
     \caption{\label{fig:Figure4} Extreme-value statistics on the spin and flux gaps. The black solid lines in (a), (c), and (d) represent the fitting of Fr\'echet distribution (Eq.~\ref{Eq:Frechet}). For the fitting curve of IDFP with $\psi = 1/2$ in (b), the approximate form in Ref. \cite{Lin2017} is used. All data are obtained from the numerical SDRG of $10^{5}$-$10^{6}$ random samples with various system sizes.}
\end{figure*}

\textit{Extreme-value statistics}.---We first consider the Ising-limit result. Based on the analytical solution of RTIC \cite{Fisher1992,Fisher1998}, in the disordered phase the extreme-value statistics of the minimum spin gap follows the Fr\'echet distribution with $u=u_{0}\Delta E_{s}L^{z}$:
\begin{align}\label{Eq:Frechet}
P(\ln u) = \frac{1}{z}u^{1/z}\exp(-u^{1/z}),
\end{align}
where the dynamical exponent $z$ is determined~\cite{Igloi2002} by the equation $z\ln\left(1-z^{-2}\right) = -\ln g$, 
and with $g = 4$ we obtain $z \approx 1.211$. In Fig.~\ref{fig:Figure4} (a), the finite-size scaling of our numerical SDRG result shows that both spin and flux gaps follow the analytical form, indicating the universal fixed-point behavior. 

In the critical phase of RTIC, the minimum-gap distribution has a rather complicated form~\cite{Fisher1998} but the finite-size scaling can be done by noting that $-\ln \Delta E \sim L^{1/2}$ from the infinite-disorder fixed point solution. Our results show that the minimum-spin-gap distributions perfectly match this scaling law (Fig.~\ref{fig:Figure4} (b), left panel), but the minimum-flux-gap distributions of different sizes cannot collapse in the same way (Fig.~\ref{fig:Figure4} (b), inset of the right panel). This suggests the non-universal behavior for the flux gaps at the critical point, which  can be understood as follows.

In the critical case ($g=1$), bond decimation is equally important as the site decimation. However, the former  may lead to a flux gap with $J_{y,i}$ from Eq.~(\ref{Eq:JzA}).
Thus, the local flux gap for $W_{2}$ can have the general expression
\begin{align}
\Delta E_{f}(W_{2}) = 2J_{y,2}+\mathrm{renormalized\,terms}\ldots
\end{align}
such that the leading term is determined by the initial distribution of $J_{y}$. This explains why in Fig.~\ref{fig:Figure3} it has linear dependence of $\Delta_{y}$ but in Fig.~\ref{fig:Figure4} (b) a non-universal size scaling is observed. To verify this, we show that the non-universal behavior in extreme-value statistics on flux gaps can be removed by defining $\Delta \tilde{E}_{f}(W_{i}) \equiv \Delta E_{f}(W_{i})/J_{y,i}$. The resulting curves (Fig.~\ref{fig:Figure4} (b), right panel) are indeed scaled as the infinite-disorder fixed-point solution.

In the XX limit, with the dominant site decimation at low energies at and away from $g = 1$, the extreme-value statistics of the spin gaps simply give the Fr\'echet distribution for uniform distributions, as shown on the left panel of Fig.~\ref{fig:Figure4} (c) and (d)
\footnote{ 
Even though the dynamical exponent is only slightly dependent on $g$ in the case of uniform distribution, it can be notably changed by using different initial distributions, such as power-law form \cite{Kaounpublished}.
This is due to the fact that the flow of energy scale is determined by $J_{z}$ which is not drastically renormalized.}.
 On the other hand, the flux-gap distributions show a trend of broadening, similar to the IDFP, but with a different exponent $\psi = 1$. Even in the case of power-law initial distributions, this scaling behavior remains robust~\cite{Kaounpublished}. The above results demonstrate again that different strong-disorder criticalities can happen for the flux excitations.

\textit{Conclusion}.---In the pursuit of quantum spin liquid phase in real materials, it was argued that disorder-induced random-single phase is  responsible for the observed power-law divergence in magnetic specific heat and susceptibility \cite{kimchi2018scaling}. In this work we showed that similar physics can emerge from the bond-anisotropic Ising systems that is relevant to the Kitaev physics, and the strong-disorder criticality appears not only for the spins but also for the local conserved quantities. 
In particular,  we studied the random Kitaev spin ladder by explicitly applying the SDRG approach.
This model
is a  quasi-one-dimensional minimal model in which spin and flux degrees of freedom coexist, and thus it provides a unique playground to study the fate of fluxes in the presence of disorder. Our findings showed that
the flux degrees of freedom in the Kitaev spin ladder survive even in the strong-disorder limit, and may reveal different criticalities compared to the spin degrees of freedom. Moreover, we found that  the power-law exponent of the flux-gap distribution depends on the type of initial distribution of couplings. This points out a new complexity in understanding the strong-disorder effect in frustrated spin systems and worth further theoretical inquiries.
Finally, since the flux-gap energy scale is usually far below the spin gaps, we anticipate that the transition between different power laws in the flux sector can be seen in the low-temperature of the specific heat  in the disordered Kitaev candidate materials \cite{Kitagawa2018spin}.


We thank Yu-Ping Lin, Eduardo Miranda, and Chi-Yun Lin for fruitful discussions. W.-H. Kao and N. B. Perkins acknowledges the support from NSF DMR-1929311 and  the support of the Minnesota Supercomputing Institute (MSI) at the University of Minnesota.
 

\bibliography{SDRG_refs.bib}
\end{document}